\documentclass[
superscriptaddress,
amsmath,amssymb,
prc,
floatfix, ]%
{revtex4-1}
\usepackage{color}
\usepackage{graphicx}
\usepackage{dcolumn}
\usepackage{bm}

\begin{document}


\title{Possible existence of bound nuclei beyond neutron drip lines driven by deformation}

\author{Xiao-Tao He}
\email{hext@nuaa.edu.cn}
 \affiliation{College of Materials Science and Technology, Nanjing University of
              Aeronautics and Astronautics, Nanjing 210016, China}
\author{Chen Wang}
 \affiliation{College of Materials Science and Technology, Nanjing University of
              Aeronautics and Astronautics, Nanjing 210016, China}
\author{Kai-Yuan Zhang}
\email{zhangky@pku.edu.cn}
\affiliation{State Key Laboratory of Nuclear Physics and Technology,
             School of Physics, Peking University, Beijing 100871, China}
\author{Cai-Wan Shen}
\email{cwshen@zjhu.edu.cn}
\affiliation{School of Science, Huzhou University, Huzhou 313000, China}
\date{\today}

\begin{abstract}
Based on the relativistic calculations of the nuclear masses in the transfermium region from No $(Z=102)$ to Ds $(Z=110)$ by the deformed relativistic Hartree-Bogoliubov theory in continuum, the possible existence of the bound nuclei beyond the neutron drip lines is studied. The two-neutron and multi-neutron emission bound nuclei beyond the primary neutron drip line of $N=258$ are predicted in $Z=106,108$ and $110$ isotopes. Detailed microscopic mechanism investigation reveals that nuclear deformation plays a vital role in the existence of the bound nuclei beyond the drip line. Furthermore, not only the quadrupole deformation $\beta_{2}$, but also the higher orders of deformation are indispensible in the reliable description of the phenomenon of the reentrant binding. 
\end{abstract}

\maketitle

\section{Introduction}

The limit of nuclear existence has been a longstanding fundamental question in nuclear science. There are about 3200 isotopes that have been confirmed experimentally so far~\cite{WangM2021_CPC45_30003,ThoennessenM2016}. Nuclear existence ends at the drip lines where there is no enough binding energy to prevent the last nucleon(s) escaping from the nucleus. On the proton-rich side, since it is closer to the stable nuclides, the drip line is relatively easy to be determined and most recently that of neptunium $(Z=93)$ has been reached experimentally~\cite{ZhangZ2019_PRL122_192503}. On the neutron-rich side, however, due to the long distance separating from the valley of stability, the neutron drip line is established experimentally only for light nuclei, i.e. from hydrogen $(Z=1)$ to neon $(Z=10)$~\cite{NeufcourtL2019_PRL122_62502}. For the heavier elements, the neutron drip line is based on the theoretical prediction and it is uncertain. 

The location of the neutron drip line and the masses of the neutron-rich nuclei are essential to understand the structure of weakly bound nuclei, the astrophysical rapid neutron capture process and the origin of elements in the universe. Exploration of very neutron-rich nuclei is extremely challenging both for nuclear theory and experiment. The production rates of the neutron-rich nuclei are very low in the experiment. It is anticipated that the development of the next generation of radioactive ion-beam will open up a new window to nuclei that were heretofore inaccessible. Theoretically, the description of weakly bound exotic systems requires proper treatment of the pairing, deformation, the continuum effects and the coupling among them.  

Beyond the neutron drip line, the phenomenon of reentrant binding is predicted in microscopic calculations~\cite{StoitsovM2000_PRC61_34311,StoitsovM2003_PRC68_54312,DelarocheJ2010_PRC81_14303,GorielyS2010_PRC82_35804,ErlerJ2012_N486_509,AfanasjevA2013_PLB726,ZhangY2013_PRC88_54305}. This phenomenon is earlier indicated in the calculations of light nuclei~\cite{StoitsovM2000_PRC61_34311} and is now spread throughout the nuclear landscape on the neutron-rich side. Predicted by various density functional theories, nuclei around $Z=60\sim100$ are found likely exist beyond the primary neutron drip lines~\cite{StoitsovM2003_PRC68_54312,DelarocheJ2010_PRC81_14303,GorielyS2010_PRC82_35804,ErlerJ2012_N486_509,AfanasjevA2013_PLB726,ZhangY2013_PRC88_54305}. Such behavior is due to the presence of shell effects at neutron closures, and local changes of the shell structure induced by deformation variations play also an important role. 

To study the loosely bound system around the neutron drip line, nuclear theory should be able to treat the asymptotic behavior of the nuclear many-body wavefunctions properly. In the deformed relativistic Hartree-Bogoliubov theory in continuum (DRHBc) theory~\cite{ZhouS2010_PRC82_11301,LiL2012_PRC85_24312}, the deformed relativistic Hartree-Bogoliubov formalism are solved in the Dirac Woods-Saxon basis in which the radial wave functions have a proper asymptotic behavior at large distance from the nuclear center~\cite{ZhouS2003_PRC68_34323}. Most recently, a new nuclear mass table is under preparation by the DRHBc theory~\cite{ZhangK2020_PRC102_24314}, which has been previously applied successfully to describe various nuclear properties~\cite{MengJ1998_PRL80_460,ZhangS2003_SiCSGM&AiC33_289,ZhangW2005_NPA753_106,MengJ2006_PPNP57_470,MengJ2015_JoPGNaPP42_93101,SunX2018_PLB785_530,SunX2020_NPA1003_122011,InE2021_IJoMPE30_2150009,YangZ2021_PRL126_82501,Sun2021_arXiv2103.10886}.

The limits of nuclear charge and mass are marked by the superheavy nuclei. It is encouraging to explore the neutron drip line and the properties of the neutron-rich nuclei in the superheavy nuclear mass region. Up to date, the heaviest nucleus observed in the experiment is the nucleus with $Z=118$ and $A =294$~\cite{OganessianY2006_PRC74_44602,HofmannS2015_JPGNPP42}. The production cross-section of superheavy nuclei is extremely low and their structural information can be rarely revealed in the experiment. An indirect way is to study lighter nuclei in the transfermium mass region with $Z>100$, which are the heaviest system accessible in present in-beam experiment~\cite{HerzbergR2008_PPNP61,HerzbergR2004_JoPGNaPP30,LeinoM2004_ARNPS54_175}. There is a hope that the ground-state properties and the nuclear structure studies of the transfermium nuclei could lead to a more reliable prediction of the location of the spherical island of stability. 

In the present work, nuclear masses near the neutron drip line from No $(Z=102)$ to Ds $(Z=110)$ are calculated by the deformed relativistic Hartree-Bogoliubov theory in continuum with density functional PC-PK1~\cite{ZhaoP2010_PRC82_54319}. The previously predicted neutron shell closure at $N=258$~\cite{ErlerJ2012_N486_509,ZhangW2005_NPA753_106} is confirmed and the possible existence of the bound nuclei beyond the neutron drip line is found in $Z=106,108$ and $110$ isotopes. The microscopic mechanism resulting in such phenomenon is analyzed. The paper is organized as follows. The DRHBc formalism and the numerical details of the calculations are briefly introduced in Sec.~\ref{sec:formalism}.  The prediction of the possible nuclear existence beyond the neutron drip lines is presented in Sec.~\ref{sec:SBNDL} and the microscopic mechanism leading to this phenomenon as well as the influence of higher orders of deformation are analyzed in Sec.~\ref{sec:mechanism}. Finally, a brief summary is given in Sec.~\ref{sec:summary}.

\section{Theoretical formalism and numerical details}
\label{sec:formalism}
The details of the DRHBc theory can be found in Refs.~\cite{LiL2012_PRC85_24312,ZhouS2010_PRC82_11301,ChenY2012_PRC85_67301,LiL2012_CPL29_42101}. The DRHBc theory with point-coupling density functional has been developed in Ref.~\cite{ZhangK2020_PRC102_24314}. For simplicity, we only give briefly the main formalism here. The relativistic Hartree-Bogoliubov equation is
\begin{eqnarray}
\label{eq:RHB}
\begin{pmatrix}
     h_{D}-\lambda_{\tau}  &  \Delta  \\
    -\Delta^{\ast}   &  -h_{D}^{\ast}+\lambda_{\tau}
\end{pmatrix}
\begin{pmatrix}
     U_{k}  \\
     V_{k} 
\end{pmatrix}
=E_{k}\begin{pmatrix}
     U_{k}  \\
     V_{k} 
\end{pmatrix},
\end{eqnarray}
which is solved in a Dirac Woods-Saxon basis~\cite{ZhouS2003_PRC68_34323}.  $E_{k}$ is the quasiparticle energy and $(U_{k} ,V_{k})^{\mathrm{T}}$ is the quasiparticle wave function. $\lambda_{\tau}$ ($\tau=n\ \mathrm{or}\ p$) is the Fermi energy. The pairing potential $\Delta$ is,
\begin{eqnarray}
\label{eq:pairing}
\Delta(\boldsymbol{r}_{1},\boldsymbol{r}_{2})=V^{pp}(\boldsymbol{r}_{1},\boldsymbol{r}_{2})\kappa(\boldsymbol{r}_{1},\boldsymbol{r}_{2}),
\end{eqnarray}
where $\kappa(\boldsymbol{r}_{1},\boldsymbol{r}_{2})$ is the pairing tensor and $V^{pp}(\boldsymbol{r}_{1},\boldsymbol{r}_{2})$ is a density dependent force of zero-range
\begin{eqnarray}
V^{pp}(\boldsymbol{r}_{1},\boldsymbol{r}_{2})=V_{0}\frac{1}{2}(1-P^{\sigma})\delta(\boldsymbol{r}_{1}-\boldsymbol{r}_{2})\left[1-\frac{\rho(\boldsymbol{r}_{1})}{\rho_{\mathrm{sat}}}\right],
\end{eqnarray}
where $V_0$ is the pairing strength, $\rho_{\mathrm{sat}}$ is the saturation density of nuclear matter, $\frac{1}{2}(1-P^{\sigma})$ is the projector for the spin $S=0$ component. 
The Dirac Hamiltonian $h_{D}$ is expressed as, 
\begin{eqnarray}
\label{eq:H}
h_{D}=\boldsymbol{\alpha}\cdot\boldsymbol{p}+V(\boldsymbol{r})+\beta[M+S(\boldsymbol{r})]
\end{eqnarray}
with $S(\boldsymbol{r})$ and $V(\boldsymbol{r})$ being the scalar and vector potentials, respectively. For an axially deformed nuclei with the reflection symmetric shape, the potential can be expanded in terms of Legendre polynomials,
\begin{eqnarray}
\label{eq:f}
f(\boldsymbol{r})=\sum_{\lambda}f_{\lambda}(r)P_{\lambda}(\cos\theta),\ \lambda=0,2,4,6\cdots
\end{eqnarray}
with
\begin{eqnarray}
\label{eq:fr}
f_{\lambda}(r)=\frac{2\lambda+1}{4\pi}\int d\Omega f(\boldsymbol{r})P_{\lambda}(\cos\theta).
\end{eqnarray}

The present calculations are carried out with the relativistic density functional PC-PK1. The numerical details are the same as that used in Ref.~\cite{ZhangK2020_PRC102_24314}. The relativistic Hartree-Bogoliubov equations are solved in a spherical Dirac Woods-Saxon basis with the box size $R_{\textrm{box}}=20$ fm and the mesh size $\Delta r=0.1$ fm. The number of states in the Dirac sea and Fermi sea is set to the same. The energy cutoff for Woods-Saxon basis is taken as $E^{+}_{\textrm{cut}} = 300$ MeV and the angular momentum cutoff is $J_{\textrm{max}} = 23/2\hbar$. The density-dependent zero-range pairing force is adopted and the pairing strength is determined by the experimental odd-even differences in binding energies. For the particle-particle channel, with a pairing window of $100$ MeV, the saturation density is $\rho_{\textrm{sat}} = 0.152$ fm$^{-3}$ and pairing strength $V_0 = -325.0$ MeV fm$^3$. We only consider the reflection-symmetric nuclear shape, i.e., $\lambda$ in Eq.~(\ref{eq:f}) is restricted to even numbers. The convergence check of the Legendre expansion has been performed~\cite{PanC2019_IJMPE29_1950082}, and $\lambda_{\textrm{max}} = 10$ is used in the present work.

\section{Possible existence of bound nuclei beyond the neutron drip line}
\label{sec:SBNDL}

\begin{figure}[ht] 
\centering
    \includegraphics[width=14cm]{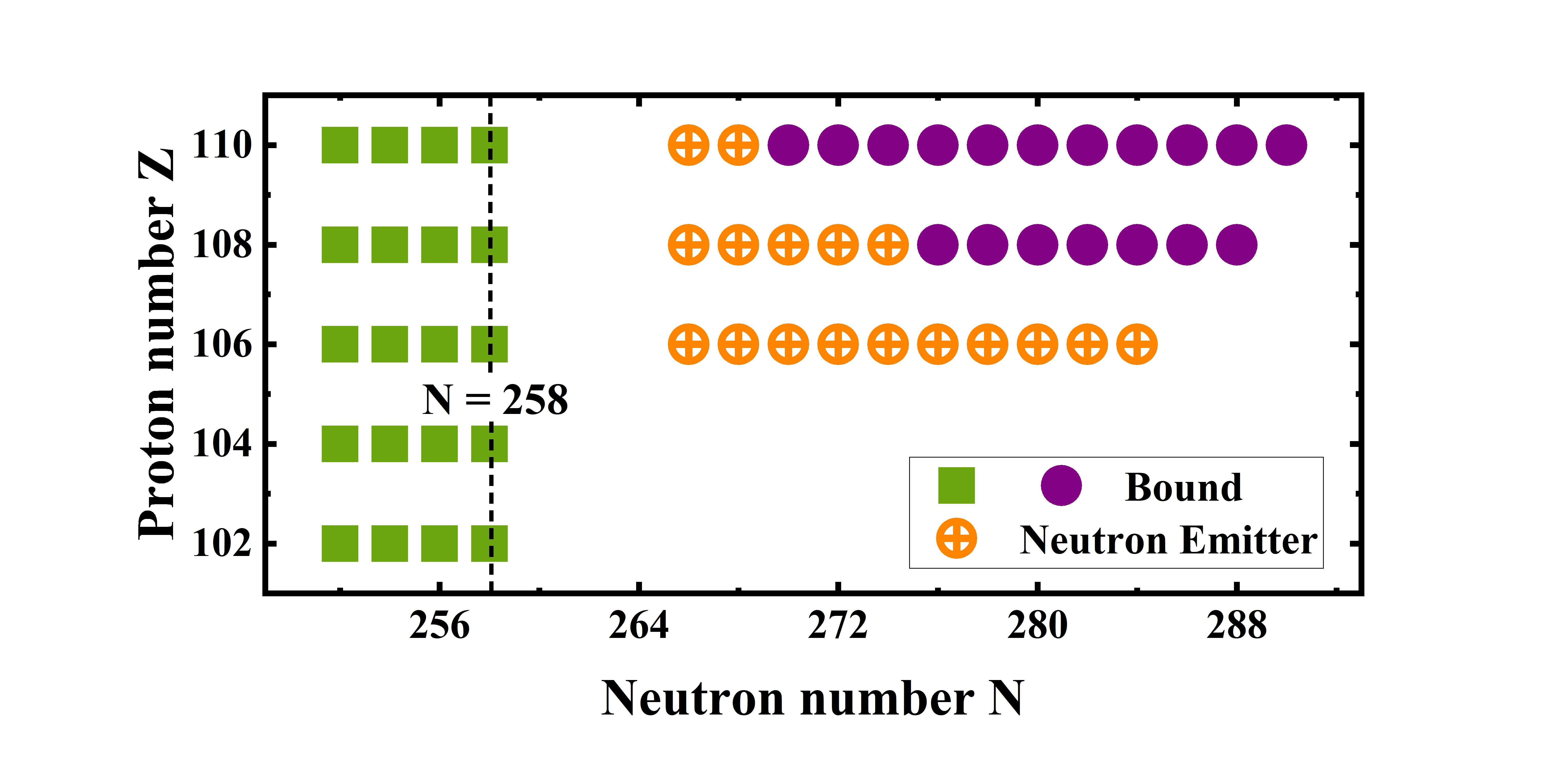}
      \caption{Section of the nuclear chart showing the possible nuclear existence beyond the neutron drip line from No $(Z=102)$ to Ds $(Z=110)$ in the DRHBc calculations. The predicted spherical neutron shell closure at $N=258$ is denoted by dashed line. The bound nuclei within and beyond the neutron drip lines are denoted by the solid squares and circles, respectively. The nuclei bound against two-neutron emission while unbound against multi-neutron emission are denoted by the crossed circles.}
\label{fig:1}
\end{figure}
 
Figure~\ref{fig:1} illustrates the section of the nuclear chart which shows the possible existence of nuclei beyond the neutron drip lines from No to Ds obtained in the DRHBc calculations. The drip line of an isotopic chain is defined by its proton/neutron separation energies. Since the present calculations are performed only for even-even nuclei, and also due to the pairing correlations, the two-neutron separation energies $S_{2n}(Z,N)=B(Z,N-2)-B(Z,N)$ are used to specify the neutron drip line, where $B(Z,N)$ is the binding energy of a nucleus with proton number $Z$ and neutron number $N$. A nucleus being bound against two-neutron emission necessitates its two-neutron separation energy $S_{2n}$ positive and the nucleus is unbound against two-neutron emission when its $S_{2n}$ is negative. 

As shown in Fig.~\ref{fig:1}, the spherical neutron closure is predicted at $N=258$ which is consistent with calculations by other density functional theories~\cite{ErlerJ2012_N486_509,AfanasjevA2013_PLB726,ThoennessenM2013_RPP76_56301}. The neutron drip line from No to Ds locate at $N=258$, and the nuclear binding ends at $N=258$ for No and Rf $(Z=104)$. It is very interesting that the phenomenon of reentrant binding beyond the neutron drip lines happens for Sg $(Z=106)$, Hs $(Z=108)$ and Ds. For Sg, nuclei with $266\leqslant N\leqslant284$ are bound against two-neutron emission, i.e. they have positive two-neutron separation energies, while unbound against multi-neutron emission, i.e. multi-neutron separation energies $B(Z,N)-B(Z=106,N=258)$ are negative. For Hs, nuclei with $266\leqslant N\leqslant288$ are bound against two-neutron emission and among these, nuclei with $266 \leqslant N \leqslant 274$ are unbound against multi-neutron emission. For Ds, the situation is similar to Hs isotopes, nuclei with $266\leqslant N\leqslant290$ are bound against two-neutron emission and unbound against multi-neutron emission for $N=266,268$.  An enhanced stability trend is demonstrated from No to Ds. It is necessary to investigate the heavier elements along this line~\cite{ZhangK2021}. 

\section{Microscopic mechanism analysis}
\label{sec:mechanism}

\begin{table*}
\centering
\caption{Ground-state properties of Ds isotopes with $252\leqslant N\leqslant294$ calculated by the DRHBc theory. The two-neutron unbound nuclei are underlined.}
\begin{ruledtabular}
\begin{tabular}{ccccccccc}
$A$  & $N$ & $B$ (MeV) &	$S_{2n}$ (MeV) & $\lambda_{n}$(MeV) & $\beta_{2}$  \\
	 \hline
$Z=110$ (Ds) \\
362	&252&      2284.175 &	3.311 &	-1.468 &	-0.084 \\
364	&254&	2287.042 &       3.311 &	-1.570 &	-0.051 \\
366	&256&	2290.220 &	3.178 &	-1.421 &	-0.034 \\
368	&258&	2293.352 &	3.132 &	-0.834 &	0.000 \\
 \underline{370}	& \underline{260}&      2292.776 &      \underline{-0.576} &	0.255 &	0.000 \\
 \underline{372}	& \underline{262}&      2292.224 &	\underline{-0.552} &	0.237 &	0.000 \\
 \underline{374}	& \underline{264}&      2291.931 &	\underline{-0.293} &	-0.113 &	0.064 \\
376	&266&	2292.448 &	0.517 &	-0.325 &	0.097 \\
378	&268&	2293.115 &	0.667 &	-0.337 &	0.118 \\
380	&270&	2293.724 &	0.609 &	-0.317 &	0.136 \\
382	&272&	2294.359 &	0.635 &	-0.358 &	0.155 \\
384	&274	&      2295.126 &	0.767 &	-0.454 &	0.177 \\
386	&276	&      2296.139 &	1.013 &	-0.555 &	0.201 \\
388	&278	&      2297.258 &	1.119 &	-0.555 &	0.218 \\
390	&280	&      2298.299 &	1.041 &	-0.481 &	0.232 \\
392	&282	&      2299.146 &	0.847 &	-0.405 &	0.243 \\
394	&284&	2299.886 &	0.741 &	-0.370 &	0.253 \\
396	&286	&      2300.566 &	0.679 &	-0.330 &	0.260 \\
398	&288	&      2301.156 &	0.590 &	-0.275 &	0.265 \\
400	&290	&      2301.593 &	0.437 &	-0.141 &	0.269 \\
\underline{402}	&\underline{292}	&      2301.674 &	0.081 &	\underline{0.008} &	0.273 \\
\underline{404}	&\underline{294}&	2301.555 &	\underline{-0.119} &	0.071 &	0.277 	
\end{tabular}
\end{ruledtabular}
\label{tab:1}
\end{table*} 

\begin{figure}[ht] 
\centering
    \includegraphics[width=16cm]{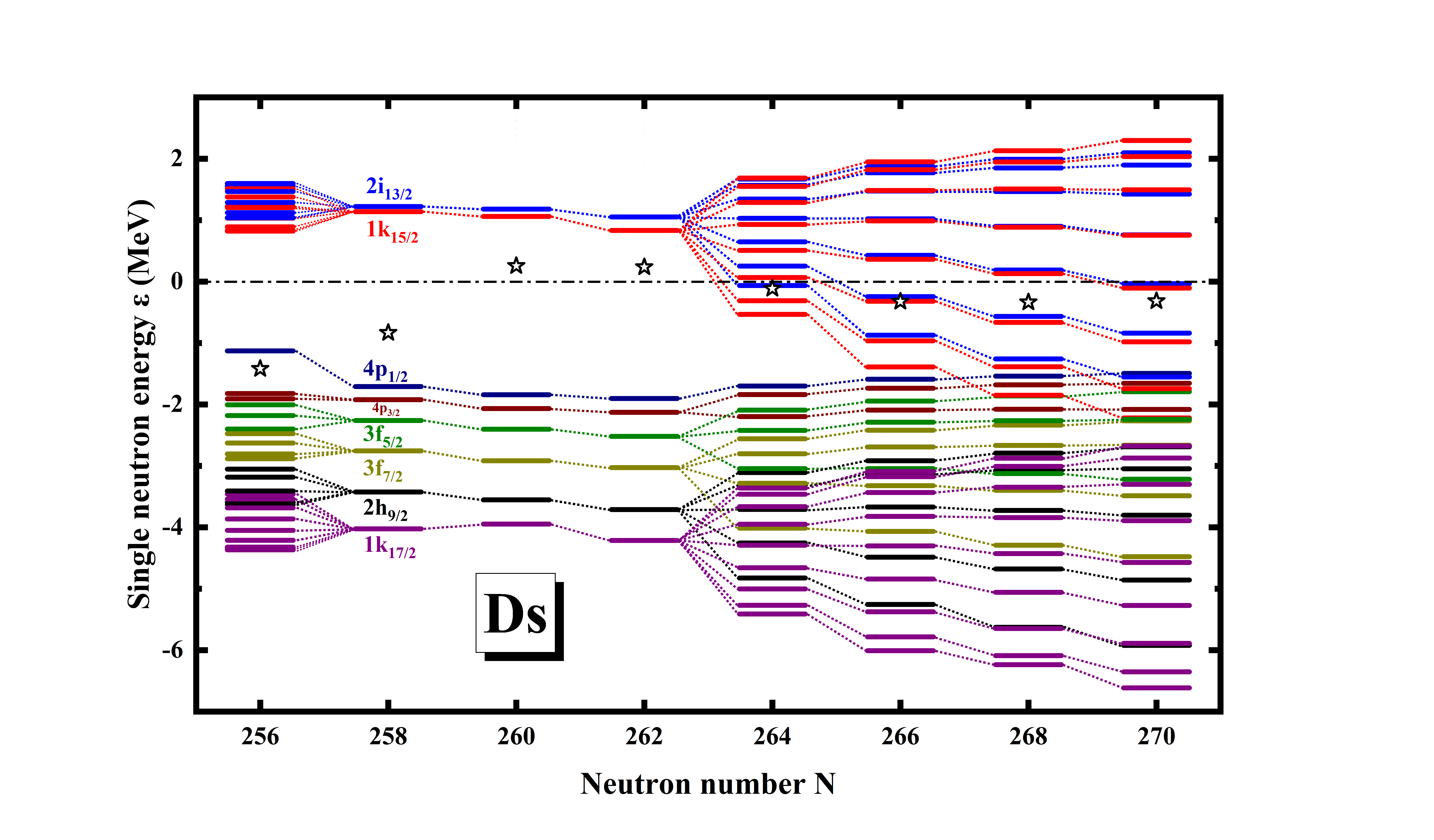}
      \caption{Single-neutron energy levels in the canonical basis near the Fermi surface of Ds isotopes with $256\leqslant N\leqslant270$. The neutron Fermi energy $\lambda_{n}$ of each isotope is denoted by hollow star.}
\label{fig:2}
\end{figure}

We take Ds isotopes near the neutron drip line as an example to reveal the microscopic mechanism of the reentrant binding beyond the drip line. Table~\ref{tab:1} shows the detailed results from $^{362}$Ds to $^{404}$Ds. The single neutron energy levels of nuclei that are just lighter and heavier than $^{368}$Ds are shown in Fig.~\ref{fig:2}. One can see that for the spherical nuclei $^{368}$Ds, $^{370}$Ds and $^{372}$Ds, a big shell gap appears for these three nuclei at $N=258$, which indicates the nature of spherical shell closure. According to the two-neutron separation energies, $^{368}$Ds is a bound nucleus whereas $^{370}$Ds and $^{372}$Ds are not. This is because the neutron Fermi surface of $^{368}$Ds is just at $N=258$ spherical shell. For the ground state, all the 258 neutrons occupy the single-particle levels below the continuum threshold (shown by the dot-dashed line in Fig.~\ref{fig:2}). For $^{370}$Ds ($^{372}$Ds), there are two (four) neutrons mainly occupying the $1k_{15/2}$ orbital which are just above the continuum threshold. Therefore, the Fermi energies of $^{370,372}$Ds are positive and their two-neutron separation energies are negative, and thus $^{370,372}$Ds are unbound against two-neutron emission. For $^{374}$Ds, six more neutrons, with comparison of $^{368}$Ds, lead to deformation of $\beta_{2}=0.064$. Three deformed single-particle levels stemming from the spherical orbitals $2i_{13/2}$ and $1k_{15/2}$ go down below the continuum threshold. A much smaller shell gap can be seen at $N=258$. For the ground state, the six neutrons mainly occupy the three levels just below the continuum threshold, and the Fermi energy becomes negative with quite small absolute value. However, $^{374}$Ds is still unbound against two-neutron emission due to the negative two-neutron separation energy, which can be explained by a loss of pairing energy. For $^{376}$Ds, the deformation increases, five deformed single-particle levels go down below the continuum threshold, and there is no energy gap around the Fermi surface. The two-neutron separation energy is positive and Fermi energy is negative. $^{376}$Ds is a two-neutron emission bound but multi-neutron emission unbound nucleus. $^{378}$Ds is very similar to $^{376}$Ds. For the heavier isotopes up to $^{400}$Ds, the deformation $\beta_{2}$ keeps increasing and more deformed single-particle levels stemming from the $2i_{13/2}$ and $1k_{15/2}$ orbitals intruder below the continuum threshold.  $^{380-400}$Ds are all bound nuclei. Therefore, we see that the nuclear deformation can strongly influence the single-particle levels and the shell structures. It plays a vital role in the reentrant binding beyond the drip lines on the presence of shell effects at neutron closure $(N=258)$.

\begin{figure}[ht] 
\centering
    \includegraphics[width=14cm]{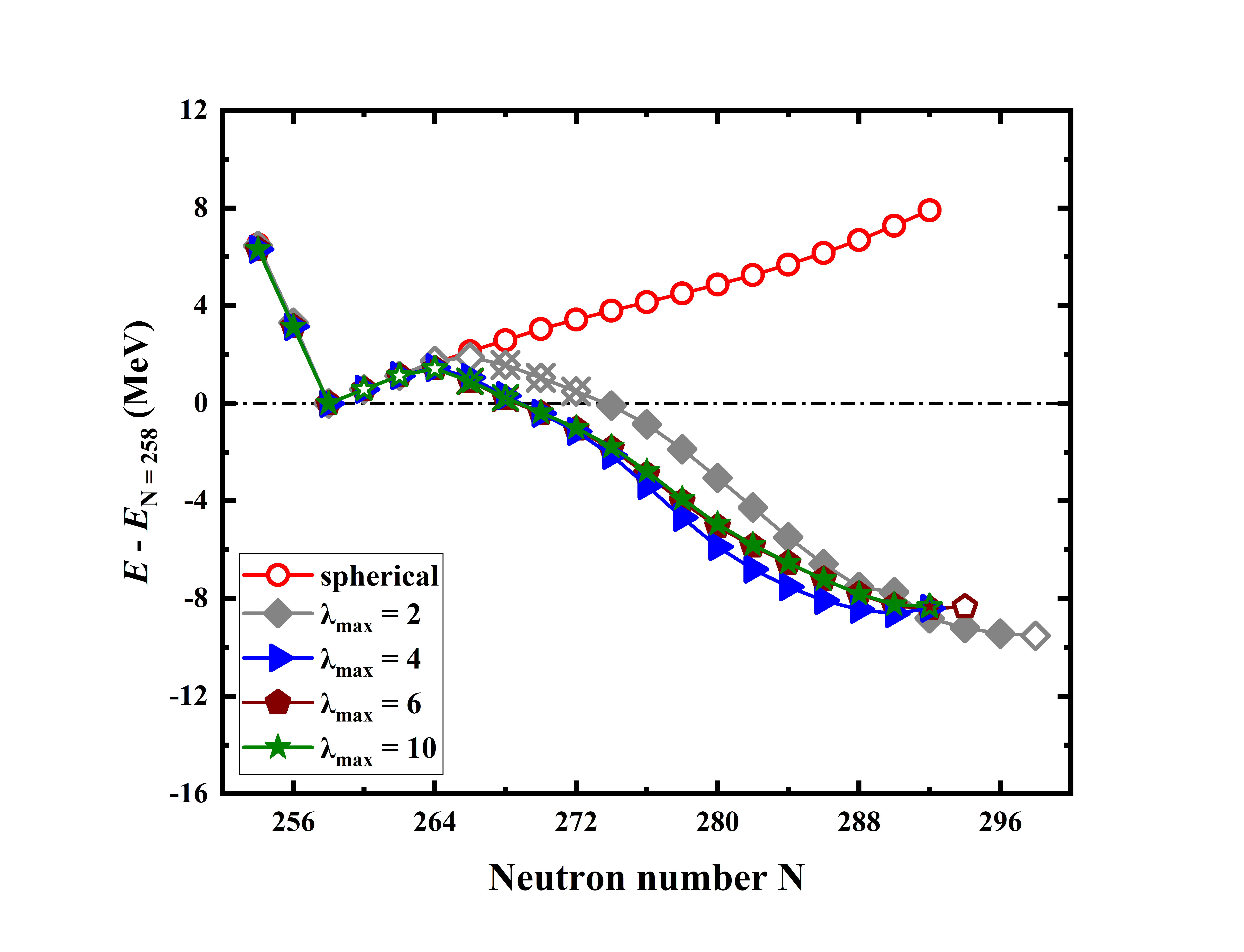}
      \caption{The DRHBc calculated total energies relative to that of $^{358}$Ds (N = 258) under different expansion order of deformation. The bound and unbound nuclei are denoted by the solid and open symbols, respectively. The nuclei which are bound against two-neutron emission while unbound against multi-neutron emission are denoted by the crossed open symbols.}
\label{fig:3}
\end{figure}

As the existence of the bound nuclei beyond the drip line is affected strongly by the nuclear deformation, it is essential to investigate how strong the influence of the different orders of deformation is on this phenomenon. We perform the DRHBc constrained calculations at $\beta_{2}=0$ and unconstrained calculations with the Legendre expansion truncation $\lambda_{\mathrm{max}} = 2$ (only including quadrupole deformation $\beta_{2}$), $\lambda_{\mathrm{max}} = 4$ (including quadrupole and hexadecapole deformations $\beta_{2}$ and $\beta_{4}$) and $\lambda_{\mathrm{max}} = 6$ (including $\beta_2$, $\beta_4$ and $\beta_6$), respectively. The calculated total energies relative to that of $^{368}$Ds $(N=258)$ as a function of the neutron number for $^{364-402}$Ds are presented in Fig.~\ref{fig:3}. One can see that for the calculations with $\beta_{2}=0$, $E-E_{N=258}$ increases at $N>258$. There would be no bound nuclei exist beyond the drip line. For the calculation of $\lambda_{\mathrm{max}} = 2$, there are 4 unbound, 3 two-neutron emission bound while multi-neutron emission unbound nuclei and 12 bound nuclei beyond the neutron drip line. Comparing with the result of $\lambda_{\mathrm{max}} = 10$, there is 1 more two-neutron emission bound but multi-neutron emission unbound nucleus beyond the drip line. For the calculation of $\lambda_{\mathrm{max}} = 4$, besides the quantitive differences, the obtained bound nuclei are as same as that of $\lambda_{\mathrm{max}} = 10$ calculation. For the calculations of $\lambda_{\mathrm{max}} = 6$, the results are very similar with that of $\lambda_{\mathrm{max}} = 10$ except that there is 1 more bound nucleus obtained in the calculations with $\lambda_{\mathrm{max}} = 6$. It demonstrates clearly that the nuclear deformation is vital to the existence of the bound nuclei beyond the drip lines. Moreover, the higher-order deformation effect is very important. 

\section{summary}
\label{sec:summary}
The deformed relativistic Hartree-Bogoliubov theory in continuum is used to calculate the nuclear masses in the transfermium region from No to Ds. The calculations lead to the prediction of the possible existence of the bound nuclei beyond the neutron drip lines in $Z=106,108$ and $110$ isotopes. We take the Ds isotopes as an example to analyze the microscopic mechanism of the phenomenon of the reentrant binding. The detailed investigations are performed based on the single-particle level structure and the occupation of the valence neutrons on the single-particle orbitals near the Fermi surface. It is found that, by locally changing the single-particle structures on the presence of shell effects at neutron closure at $N=258$, the nuclear deformation plays a vital role in the existence of the bound nuclei beyond the neutron drip lines. Further investigation shows that not only the quadrupole deformation $\beta_{2}$, but also the hexadecapole deformation $\beta_{4}$ and high-order deformation $\beta_{6}$ are indispensible in the reentrant binding beyond the drip line.

\begin{acknowledgements}
We thank the members of DRHBc Mass Table Collaboration for helpful discussions. This work is supported by the National Natural Science Foundation of China (Grant Nos. U2032138, 11775112, 12075085, 11935003 and 12047568) and the State Key Laboratory of Nuclear Physics and Technology, Peking University (Grant No. NPT2020ZZ01). 
\end{acknowledgements}


\bibliographystyle{apsrev4-1}
\bibliography{../../../References/ReferencesXT}

\end{document}